\RequirePackage{fix-cm}
\documentclass[smallextended]{svjour3}       % onecolumn (second format)
\smartqed  % flush right qed marks, e.g. at end of proof

\usepackage{amsmath}
\usepackage{graphicx}
\usepackage{amssymb}
\usepackage{amsfonts}
\usepackage{graphicx,color}
\usepackage{amsbsy}

\usepackage[portuguese]{babel}
\begin{document}

\newtheorem{Def}{Definition}[section]
\newtheorem{Proc}{Procedure}[section]
\newtheorem{Lemma}{Lemma}[section]
\newtheorem{Thm}{Theorem}[section]
\newtheorem{Corollary}{Corollary}[section]
\newtheorem{Problem}{Problem Statement}[section]
\newtheorem{Alg}{Algorithm}[section]
\newtheorem{Opm}{Remark}[section]
\newtheorem{Ex}{Example}[section]
\newtheorem{Property}{Property}[section]
\newtheorem{Remark}{Remark}[section]
\newtheorem{Proof}{Proof}

\newcommand{\AD}{\mathcal{A}_{p}[D]}
\newcommand{\A}{\mathcal{A}_{p}}

\newcommand{\ZZ}{\Z_{p^r}}
\newcommand{\ZZD}{\Z_{p^r}[D]}
\newcommand{\ellL}{L}
\newcommand{\kpar}{k}
\newcommand{\qpar}{q}
\newcommand{\dubbel}[1]{{\mathbb #1}}
\newcommand{\F}{{\mathbb F}}
\newcommand{\Z}{{\mathbb Z}}
\newcommand{\es}{x}
\newcommand{\st}{ \;|\;}
\newcommand{\ruimte}{\par\vspace{1ex}\noindent}
\newcommand{\ruimtevier}{\par\vspace{4ex}\noindent}
\newcommand{\Fpolstimes}[2]{{\mathbb F}^{#1 \times #2} [s]}
\newcommand{\Rpolstimes}[2]{{\mathbb R}^{#1 \times #2} [s]}
\newcommand{\wt}{\mathrm{wt}}
\newcommand{\et}{\widetilde{\ebold}}
\newcommand{\suitc}{s \in \mC}
\newcommand{\keenN}{k = 1,2,\ldots , N}
\newcommand{\knulN}{k = 0,1,\ldots , N}
\newcommand{\arij}{a_1, a_2, \ldots ,a_N}
\newcommand{\ariji}{a_1^i, a_2^i, \ldots ,a_N^i}
\newcommand{\arijK}{a_1, a_2, \ldots ,a_K}
\newcommand{\Arijnul}{A_0, A_1, \ldots ,A_N}
\newcommand{\Arij}{A_1, \ldots ,A_N}
\newcommand{\arijk}{a_1, a_2, \ldots ,a_k}
\newcommand{\arijkp}{a_1, a_2, \ldots ,a_{k+1}}
\newcommand{\arijkm}{a_1, a_2, \ldots ,a_{k-1}}
\newcommand{\crij}{c_1, c_2, \ldots ,c_L}
\newcommand{\impliess}{\;\Longrightarrow\;}
\newcommand{\pf}{\paragraph{Proof}}
\newcommand{\pfend}{\par\vspace{2ex}\noindent}
\newcommand{\eind}{\hspace*{\fill}$\Box$\par\vspace{2ex}\noindent}
\newcommand{\beq}{\begin{equation}}
\newcommand{\eeq}{\end{equation}}
\newcommand{\bmat}{\left[ \begin{array}}
\newcommand{\emat}{\end{array} \right]}
\newcommand{\twee}[2]{\left[ #1 \sfour #2 \right]}
\newcommand{\B}{{\cal B}}
\newcommand{\R}{{\cal R}}
\newcommand{\C}{{\cal C}}
\newcommand{\COL}{\mathrm{col }\;}
\newcommand{\im}{\mbox{ Im }}

\title{On MDS convolutional Codes over $\Z_{p^r}$ %\thanks{Grants or other notes
}


\author{Diego Napp \and Raquel Pinto \and Marisa Toste  }


\institute{Raquel Pinto \and
        Diego Napp \and
         \at
              Department of Mathematics, University of Aveiro, Portugal \\
 %             Tel.: +123-45-678910\\
 %             Fax: +123-45-678910\\
              \email{diego@ua.pt}           %  \\
           \and
           Marisa Toste \at
              Superior School of Technologies and Management of Olveira do Hospital, Polytechnic Institute of Coimbra, Coimbra, Portugal
}

\date{Received: date / Accepted: date}

\maketitle

\begin{abstract}
Maximum Distance Separable (MDS) convolutional codes are characterized through the property that the free distance meets the generalized Singleton bound. The existence of \emph{free} MDS convolutional codes over $\ZZ$ was recently discovered in \cite{el13} via the Hensel lift of a cyclic code. In this paper we further investigate this important class of convolutional codes over $\ZZ$ from a new perspective. We introduce the notions of p-standard form and r-optimal parameters to derive a novel upper bound of Singleton type on the free distance. Moreover, we present a constructive method for building general (non necessarily free) MDS convolutional codes over $\ZZ$ for any given set of parameters.
\end{abstract}

\section{Introduction}

The extension of the concept of convolutional codes from finite fields to finite rings was first developed in \cite{massey89} and have attracted much attention in recent years. This interest is mainly due to the discover that the most appropriate codes for phase modulation are the linear codes over the residue class ring $\Z_M$, $M$ a positive integer. It was immediately apparent that convolutional codes over rings behave very different from convolutional codes over finite fields. For instance, in contrast with the field case, (linear) convolutional codes over finite rings $\R$ are not necessarily free modules over $\R$.

Fundamental results of the structural properties of convolutional codes over finite rings can be found in, for instance, \cite{fa01,jo98,no99,no00a}. In particular, the properties of noncatastrophic, right invertible, basic and systematic ring convolutional encoders were thoroughly discussed. The problem of deriving minimal encoders (left prime and row-reduced) was posed in \cite{fa97,so07}. This problem was solved in \cite{ku07,ku09} using the concept of minimal p-encoder, which is an extension of the concept of $p$-basis introduced in \cite{va96} to the polynomial context.

In \cite{as94,ko95} the search for and design of unit-memory convolutional codes over $\Z_4$ that gives rise to binary trellis codes with high free distances was investigated and several concrete constructions were reported. In \cite{jo98a} two 16-state trellis codes of rate $\frac{2}{4}$, again over $\Z_4$, were found by computer search. However, in contrast to the block code case \cite{gu12,no01} little is known about distance properties and constructions of convolutional codes over large rings, see for instance \cite{so07}.

%In this paper we seek to further investigate the distance properties of convolutional codes over $\Z_{p^r}$, for $p$ prime. In particular we consider the most fundamental distance measure of convolutional codes, which is the column distance \cite{co69,jo99}.

Recently, in \cite{el13} a bound on the free distance of convolutional codes over $\Z_{p^r}$ was derived, generalizing the bound given in \cite{ro99a1} for convolutional codes over finite fields. Codes achieving such a bound were called Maximal Distance Separable (MDS). %Note that MDS convolutional codes exist only over sufficiently large rings (or fields).
The concrete constructions of MDS convolutional codes over $\Z_{p^r}$ presented in \cite{el13} were restricted to \emph{free} codes and cannot be extended to the general case. An explicit general construction of nonfree MDS codes over finite rings was left as an open problem.

In this paper we adopt a simple but novel approach to further investigate this important class of convolutional codes over $\ZZ$. In particular, we derive new upper-bounds on the free distance and provide explicit novel constructions of nonfree MDS convolutional codes over $\ZZ$ for every set of given parameters. In the proof of these results, an essential role is played by the theory of $p$-basis and in particular of a canonical form of the $p$-encoders. In contrast with the papers \cite{no01,el13} where the Hensel lift of a cyclic code was used, in this paper a direct lifting is employed  to build convolutional codes over $\ZZ$ from known constructions of convolutional codes over $\Z_p$. Note that by the Chinese Remainder Theorem, results on codes over $\Z_{p^r}$ can be extended to codes over $\Z_M$, see also \cite{ch94,jo98,McD74}.

The paper is organized as follows: In the next section we introduce some preliminaries on  p-basis of $\dubbel{Z}_{p^r}[D]$-submodules of $\dubbel{Z}_{p^r}^n[D]$. After presenting block codes over $\dubbel{Z}_{p^r}$ we introduce the new notions of p-standard form and r-optimal parameters. Using these notions a novel Singleton-type upper-bound  is derived. In section 3 we consider convolutional codes and provide the basic definitions. An upper bound for their free distance is presented. Finally, we propose a method to build MDS convolutional codes over $\dubbel{Z}_{p^r}$ for any given set of parameters in section 4.

\section{Preliminaries}

\subsection{$P$-basis and $p$-dimension}

Any element in $\dubbel{Z}_{p^r}^n$ can be written uniquely as a linear combination of $1,p,p^2,\dots$ $ \dots, p^{r-1}$, with coefficients in $\mathcal{A}_p=\{0,1, \dots,p-1\} \subset \dubbel{Z}_{p^r}$ (called the $p$-adic expansion of the element) \cite{ca00a}. Note that all elements of $\mathcal{A}_p \backslash\{0\}$ are units. This set will play an important role throughout the paper since it will allow us to introduce the notion of $p$-basis of $\dubbel{Z}_{p^r}$-submodule of $\dubbel{Z}_{p^r}^n$, which will be crucial in the analysis and construction of optimal convolutional codes over $\dubbel{Z}_{p^r}$.

Let $v_1(D), \dots, v_k(D) $ be in $\dubbel{Z}^n_{p^r}[D]$. The vector $\displaystyle \sum_{j=1}^k a_j(D) v_j(D)$,
with $a_j(D) \in \mathcal{A}_p [D]$, is said to be a {\bf $\boldsymbol{p}$-linear combination} of ${v_1(D), \dots, v_k(D)}$ and the set of all $p$-linear combination of ${v_1(D), \dots, v_k(D)}$ is called the {\bf $\boldsymbol{p}$-span} of $\{v_1(D), \dots, v_k(D) \}$, denoted by
$p$-span $(v_1(D), \dots, v_k(D))$. An ordered set of vectors $(v_1(D), \dots, v_k(D))$ in $\dubbel{Z}^n_{p^r}[D]$ is said to be a {\bf $\boldsymbol{p}$-generator sequence} if
$p \, v_i(D)$ is a $p$-linear combination of $v_{i+1}(D), \dots, v_k(D)$, $\; \; i=1, \dots, k-1$,
and $p \, v_k(D)=0$.

If $(v_1(D), \dots, v_k(D))$ is a $p$-generator sequence it holds (see for instance \cite{ku07}) that
$
p\mbox{-span}(v_1(D), \dots, v_k(D))=\mbox{span}(v_1(D), \dots, v_k(D)),
$
 and consequently the  $p$-span$(v_1(D), \dots, v_k(D))$ is a $\ZZ$-submodule of $\dubbel{Z}^n_{p^r}[D]$. Note that if $M=span(v_1(D), \dots, v_k(D))$,
\begin{equation}\label{eq:07}
\begin{split}
(v_1(D), pv_1(D) \dots, & p^{r-1}v_1(D),v_2(D), pv_2(D), \dots, \\
& \dots, p^{r-1}v_2(D), \dots, v_l(D), pv_k(D) \dots, p^{r-1}v_k(D)).
\end{split}
\end{equation}
is a $p$-generator sequence of $M$.

 %If $M=p$-span$(v_1(D), \dots,v_k(D))$ we say that $(v_1(D), \dots, v_k(D))$ is a $p$-generator sequence of $M$.

The vectors $v_1(D), \dots, v_k(D)$  in $\dubbel{Z}_{p^r}^n[D]$ are said to be {\bf $\boldsymbol{p}$-linearly independent} if the only $p$-linear combination of
$v_1(D), \dots, v_k(D)$ that is equal to $0$ is the trivial one.

An ordered set of vectors $(v_1(D), \dots, v_k(D))$ which is a $p$-generator sequence of $M$ and $p$-linearly independent is said to be a {\bf $\boldsymbol{p}$-basis} of $M$. It is proved in \cite{ku09} that two $p$-bases of a $\dubbel{Z}_{p^r}$-submodule $M$ of $\dubbel{Z}_{p^r}^n[D]$ have the same number of elements. This number of elements is called {\bf $\boldsymbol{p}$-dimension} of $M$.

A nonzero polynomial vector $v(D)$ in $\dubbel Z_{p^r}^n[D]$, written as $v(D)=\sum\limits_{t=0}^{\nu}{v_tD^t}$, with $v_t \in \mathbb Z_{p^r}^n$, and $v_{\nu} \neq 0$, is said to have degree $\nu$, denoted by $deg \, v(D)=\nu$, and $v_{\nu}$ is called the {\bf leading coefficient vector} of $v(D)$, denoted by $v^{lc}$. For a given matrix $G(D)\in \dubbel{Z}_{p^r}^{k \times n}[D]$ we denote by $G^{lc} \in \dubbel{Z}_{p^r}^{k \times n}$ the matrix whose rows are constituted by the leading coefficient of the rows of $GD)$. A $p$-basis $(v_1(D), \dots, v_k(D))$ is called a {\bf reduced $\boldsymbol{p}$-basis} if the vectors $v_1^{lc}, \dots, v_k^{lc}$ are $p$-linearly independent in $\dubbel Z_{p^r}^n$. %Analogously one can define the notion of reduced basis for a submodule of $\dubbel Z_{p^r}^n[D]$.

By \cite{ku07} every submodule $M$ of $\dubbel{Z}_{p^r}^n[D]$ has a reduced $p$-basis. Note that $M$ does not always admit a (reduced) basis. Moreover, any reduced $p$-basis $(v_1(D), \dots, v_k(D))$ of $M$ exhibits the {\bf $\boldsymbol{p}$-predictable degree property} \cite{ku07}:
$$
deg \, \left(\sum\limits_{i=1}^{k}{a_i(D)v_i(D)}\right)= \max_{j:a_j(D) \in \mathcal{A}_p[D]\backslash \{0\}}(deg \, a_j(D)+deg \, v_j(D))
$$

The degrees of the vectors of  two reduced $p$-bases of $M$ are the same (up to permutation) and their sum is called the {\bf $\boldsymbol{p}$-degree} of $M$. %Moreover, there always exists a reduced $p$-basis of $M$, $(v_1(D), \dots, v_k(D))$ such that $deg \, v_i(D) \geq deg \, v_{i+1}(D)$, $i=1, \dots, k-1$.

\subsection{Block Codes}

A {\bf (linear) block code $\mathcal{C}$} of length $n$ over $\dubbel{Z}_{p^r}$ is a $\dubbel{Z}_{p^r}$-submodule of $\dubbel{Z}^n_{p^r}$ and the elements of $\C$ are called codewords. A generator matrix $\widetilde G\in \ZZ^{\widetilde k \times n}$ of $\mathcal{C}$ is a polynomial matrix whose rows form a minimal
set of generators of $\mathcal{C}$ over $\ZZ$. If $\widetilde G$ has full row rank, then it is called an encoder of $\C$ and $\C$ is a free module.
If $\mathcal{C}$ has $p$-dimension $k$,
a {\bf $\boldsymbol{p}$-encoder} $G \in \dubbel{Z}^{k \times n}_{p^r}$ of $\mathcal{C}$ is a matrix whose rows form a $p$-basis of $\mathcal{C}$ and therefore
$$
\mathcal{C}=\mbox{Im}_{\mathcal{A}_p}G=\{v=uG \in \dubbel{Z}^n_{p^r} \ : u \in \mathcal{A}^k_p\}.
$$

Next we introduce the notion of p-standard form that will play an important role in the sequel. Given a $p$-basis $(v_1, \dots, v_k)$ of $\mathcal{C}$ there are certain elementary operations that can be applied to $(v_1, \dots, v_k)$ so that we obtain another $p$-basis of $\mathcal{C}$. These are described in the following lemma which is not difficult to prove.
\begin{Lemma} \label{oper}
Let $(v_1, \dots, v_k)$ be a $p$-basis of a submodule $M$ of $\dubbel{Z}^n_{p^r}$. Then,
\begin{enumerate}
  \item If $v'_i=v_i+\sum_{j=i+1}^{k}a_jv_j$, with $a_j \in \dubbel{Z}_{p^r}$, then $(v_1, \dots, v_{i-1}, v'_i, v_{i+1}, \dots, v_k)$ is a $p$-basis of $M$;
  \item If $pv_i$ is a $p$-linear combination of $v_j,v_{j+1}, \dots, v_k$, for some $j>i$, then $(v_1, \dots, v_{i-1}, v_{i+1}, \dots, v_{j-1}, v_i, v_j, \dots, v_k)$ is a $p$-basis of $M$.
\end{enumerate}
\end{Lemma}

Given a generator matrix of $\C$ in standard form as in \cite{ca00a,no01}, it is easy to see that we can extend it as in (\ref{eq:07}) and apply the elementary row operations (as defined in Lemma \ref{oper} and deleting the zero rows) to obtain \emph{a} $p$-encoder $G$ in the following form:
{
\fontsize{7}{7}\selectfont
\arraycolsep=3pt % default: 5pt
$$
\!\!\!\!\!\! \left[\!\!\!\!
  \begin{array}{ccccccc}
    I_{k_0} & A^0_{1,0} & A^0_{2,0} & A^0_{3,0} & \cdots & A^0_{r-1,0} & A^0_{r,0} \\
    ------&------&------&------&------&------&------ \\
    pI_{k_0} & 0 & pA^0_{2,1} & pA^0_{3,1} & \cdots & pA^0_{r-1,1} & pA^0_{r,1} \\
    0 & pI_{k_1} & pA^1_{2,1} & pA^1_{3,1} & \cdots & pA^1_{r-1,1} & pA^1_{r,1} \\
    ------&------&------&------&------&------&------ \\
    p^2I_{k_0} & 0 & 0 & p^2A^0_{3,2} & \cdots & p^2A^0_{r-1,2} & p^2A^0_{r,2} \\
    0 & p^2I_{k_1} & 0 & p^2A^1_{3,2} & \cdots & p^2A^1_{r-1,2} & p^2A^1_{r,2} \\
    0 & 0 & p^2I_{k_2} & p^2A^2_{3,2} & \cdots & p^2A^2_{r-1,2} & p^2A^2_{r,2} \\
    ------&------&------&------&------&------&------ \\
    \vdots&\vdots&\vdots&\vdots&\cdots&\vdots&\vdots\\
    ------&------&------&------&------&------&------ \\
    p^{r-1}I_{k_0} & 0 & 0 & 0 & \cdots & 0 & p^{r-1}A^0_{r,r-1} \\
    0 & p^{r-1}I_{k_1} & 0 & 0 & \cdots & 0 & p^{r-1}A^1_{r,r-1} \\
    0 & 0 & p^{r-1}I_{k_2} & 0 & \cdots & 0 & p^{r-1}A^2_{r,r-1} \\
    0 & 0 & 0 & p^{r-1}I_{k_3} & \cdots & 0 & p^{r-1}A^3_{r,r-1} \\
    \vdots&\vdots&\vdots&\vdots&\ddots&\vdots&\vdots\\
    0 & 0 & 0 & 0 & \cdots & p^{r-1}I_{k_{r-1}} & p^{r-1}A^{r-1}_{r,r-1} \\
  \end{array}
\!\!\!\! \right]
$$
}where $I_\ell$ denotes the identity matrix of size $\ell$. One can verify that the scalars $k_i$, $i=0,1, \dots, r-1$, are equal for all $p$-encoders of $\C$ in this form, \emph{i.e.}, they are uniquely determined for a given code $\mathcal{C} \subset \dubbel{Z}_{p^r}^n$. We call $k_0, k_1, \dots, k_{r-1}$ the {\bf parameters} of $\mathcal{C}$. Clearly, if $\C$ has $p$-dimension equal to $k$ then $k= \sum_{i=0}^{r-1} k_i (r-i) $. If $G$ is in such a form we say that $G$ is in the {\bf $\boldsymbol{p}$-standard form}. The p-standard form will be a useful tool to prove our results in the same way the standard form was for previous results in the literature, see for instance \cite{ca00a,no01}. \\

The {\bf free distance $d(\mathcal{C})$} of a linear block code $\mathcal{C}$ is given by
$$
d(\mathcal{C})=\min\{\wt (v), v \in \mathcal{C}, v\neq 0 \}
$$
where $\wt (v)$ is the Hamming weight of $v$ over $\dubbel{Z}_{p^r}$.\\

%Since $\wt (v)\geq \wt (pv)$, for all $v \in \dubbel{Z}_{p^r}^n$, the following Lemma is immediately.

%\begin{Lemma}
%Let $\mathcal{C}$ be a block code over $\dubbel{Z}_{p^r}$. Then exist a $v \in p^{r-1}\dubbel{Z}_{p^r}^n$, $v \in \mathcal{C}$, such that $dist(\mathcal{C})=\wt (v).$
%\end{Lemma}

Considering the last row of any $p$-encoder in the $p$-standard form as a codeword, the next result on the generalized Singleton bound on the free distance of codes over $\ZZ$ readily follows.

\begin{Thm}\cite{no01}
Given a linear block code $\mathcal{C} \subset \dubbel{Z}_{p^r}^n$ with parameters $k_0,\dots,k_{r-1}$, it must hold that
$$
d(\mathcal{C})\leq n-(k_0 + \dots + k_{r-1})+1.
$$
\end{Thm}

Among block codes of length $n$ and $p$-dimension $k$, we are interested in the ones with largest possible distance. Hence, given an integer $r\geq 1$ and a non-negative integer $k$ we call the ordered set $(k_0, k_1, \cdots, k_{r-1})$, $k_i \in \dubbel{N}, \ i=0, \cdots, r-1$ an {\bf $\boldsymbol{r}$-optimal set of parameters} of $k$ if
$$
 k_0+k_1+ \cdots+k_{r-1} = \min_{k=rk'_0+(r-1)k'_1+ \cdots+k'_{r-1}} (k'_0+k'_1+ \cdots+k'_{r-1}).
$$

Note that when $\C$ is free, $k$ must divide $r$ and $k_0=\frac{k}{r}$, {\em i.e.}, $(k_0,0,\dots, ,0)$ is the unique $r$-optimal set of parameters of $k$. However, the $r$-optimal set of parameters of $k$ is not necessarily unique for given $k,r$. For instance if $k=25$ and $r=6$, $(4,0,0,0,0,1)$ and $(0,5,0,0,0,0)$ are two possible $6$-optimal set of parameters of $25$. The computation of the $r$-optimal set of parameters is the well-known change making problem \cite{ch70}.

\begin{Lemma}\label{opt_parameters}
Let $(k_0, k_1, \cdots, k_{r-1})$ be an $r$-optimal set of parameters of $k$. Then, $k_0+k_1+ \cdots+k_{r-1}=\left\lceil\frac{k}{r}\right \rceil$.
\end{Lemma}

\pf
Write $k=rb+a$, where $b,a \in \mathbb{N}$ and $a<r$. Note that $a$ can be written as $a=r-i$, for some $1\leq i \leq r$.
If $r|k$ then $a=0$ and necessarily $k_0=\frac{k}{r}$ and $k_j=0$, for $1 \leq j \leq r-1$.
If $r \nmid k$, we can select    $k_0=b$, $k_{r-a}=1$ and $k_j=0$, for $j \in \{1, \dots r-1 \} \backslash\{r-a\}$. Hence $k_0+k_1+ \cdots+k_{r-1}=b + 1= \left\lceil\frac{k}{r}\right \rceil$. It is easy to verify that these values  minimize $k_0+k_1+ \cdots +k_{r-1}$ subject to $k=rk_0+(r-1)k_1+ \cdots+k_{r-1}$.
\eind \pfend

Using the previous lemma the Singleton bound of codes over $\ZZ$ in terms of the $p$-dimension reads as follows.

\begin{Corollary}
Given a block code $\mathcal{C}\subset \dubbel{Z}_{p^r}^n$ and $p$-dimension $k$,
$$
d(\mathcal{C})\leq n- \left \lceil\frac{k}{r}\right \rceil +1.
$$
\end{Corollary}

Using completely different approach this result was also derived in \cite[Theorem 3.1]{el13} without using the notions of $p$-standard form nor the $r$-optimal set of parameters. Note, however, that our approach and in particular these two notions will turn out to be crucial to derive our results in the following two sections.
%%%%%%%%%%%%%%%%%%%%%%%%%

\section{Convolutional Codes}

A {\bf convolutional code $\mathcal{C}$} of length $n$ is a $\dubbel{Z}_{p^r}[D]$-submodule of  $\dubbel{Z}^n_{p^r}[D]$. A generator matrix $\widetilde G(D)\in \ZZ^{\widetilde k \times  n}[D]$ of $\mathcal{C}$ is a polynomial matrix whose rows form a minimal set of generators of $\mathcal{C}$ over $\ZZ[D]$. If $\widetilde G(D)$ has full row rank, then it is called an encoder of $\C$ and $\C$ is a free code.

If $\mathcal{C}$ has $p$-dimension $k$, a {\bf $\boldsymbol{p}$-encoder} $G(D) \in \dubbel{Z}_{p^r}^{k\times n}[D]$ of $\mathcal{C}$ is a polynomial matrix whose rows form a $p$-basis of $\mathcal{C}$ and therefore
$$
{\mathcal C}  =  \im_{\mathcal{A}_{p}[D]}G(D)
= \left\{u(D)G(D) \in \dubbel{Z}^n_{p^r}[D] :\, u(D) \in \mathcal{A}^k_p[D]\right\}.
$$

If the rows of $G(D)$ ($\widetilde G(D)$) form a reduced $p$-basis (basis) then we say the $G(D)$ ($\widetilde G(D)$) is in reduced form. The $p$-degree of $\C$, denoted by $\delta$, is the sum of the row degrees of any $p$-encoder in reduced form. In the sequel, we will adopt the notation used by McEliece \cite[p. 1082]{mc98} and denote by $(n,k,\delta)$- convolutional code a code $\mathcal{C}\subset \ZZ^n[D] $ with $p$-dimension $k$ and $p$-degree $\delta$.

\begin{remark}
 We emphasize that in this paper we do not assume that $\C$ is free. Note that convolutional codes $\C\subset \ZZ^n[D]$ always admit a $p$-encoder however they may not admit a full row rank generator matrix, {\em i.e.}, an encoder. The difference is that the input vector takes values in $\AD$ for $p$-encoders whereas for generator matrices takes values in $\ZZ [D]$. This idea of using a $p$-adic expansion for the information input vector is already present in, for instance, \cite{ca00a} and was further developed in \cite{va96} introducing the notion of $p$-generator sequence of vectors in $\ZZ$. In \cite{ku09,ku07} this notion was extended to polynomial vectors.
\end{remark}

\begin{remark}
Conform \cite{AlmeidaNappPinto2013,ku08,Napp2010,ro96a1,so07} we have decided to define our codes as \emph{finite support } convolutional codes. There exists however a considerable body of literature in which code sequences are semi-infinite Laurent series \cite{fa01,jo98,massey89,no99,no00a}. We note that for the issues treated in this paper there is no difference and all our results apply to both approaches.
\end{remark}

The  weight of $v(D)$ is given by $\wt (v(D))=\sum_{i \geq 0}{\wt (v_i)}$ and the {\bf free distance} of a convolutional code $\mathcal{C}$ is defined as
$$
d(\mathcal{C})=\min\{\wt (v(D)): \, v(D) \in \mathcal{C}, \, v(D) \neq 0\}.
$$

%%%%%%%%%%%
%\begin{Def} \label{sd}
%Let $\C$ be a $(n,k,\delta)$-convolutional code defined over an arbitrary $\ZZ$. We define the smallest degree as  and the smaller degree subcode of $\C$ as
%$$
%\C_{sd}=\{v(D) \in M:\,deg(v(D))=m \vee v(D)=0,\; m=min_{v \in M\setminus\{0\}}\,deg(v(D))\},
%$$
%and
%$$
%M_{sd}^{lc}=\{v^{lc}: v(D) \in M_{sd} \setminus \{0\}\} \cup \{0\},
%$$
%where $m$ is the smallest $p$-index of $M$ and $v^{lc}$ the leading coefficient vector of $v(D)$.
%\end{Def}
%
%
%\begin{Lemma}\label{sd}
%The sets $M_{sd}$ and $M_{sd}^{lc}$ are submodules of $\mathbb Z_{p^r}^n[D]$.
%\end{Lemma}
%
%\pf
%TO DO
%\eind \pfend

The $j$-th row distance $d^r_j$ of a $p$-encoder in reduced form $G(D)$ \cite{jo99} is defined as the minimum of the weights of all finite codewords resulting from an information sequence $u(D)\in \mathcal{A}^k_p[D]$ with $deg(u(D))\leq j$, {\em i.e.},
$$
d^r_j = \min_{deg(u(D))\leq j} \wt (u(D)G(D)).
$$
Clearly, if $\C =\im_{\mathcal{A}_{p}[D]} G(D)$,
\begin{equation}\label{eq:00}
  d(\mathcal{C})\leq \dots \leq d^r_j \leq \dots \leq d^r_1 \leq d^r_0.
\end{equation}

Let $\C$ be a $(n,k,\delta)$-convolutional code defined over $\ZZ$. Let $G(D)= G_0 + G_1 D + \cdots + G_{\nu_1}D^{\nu_1}$ be a $p$-encoder in reduced form with ordered row degrees $\nu_1 \geq \nu_2 \cdots \geq \nu_k$, and let $\nu=\min\{ \nu_1,\nu_2, \dots, \nu_k\}$ denote the value of the smallest row degree and $\ell$ the number of rows with row degree equal to $\nu$. After applying row permutation and elementary row operations we can bring the last $\ell$ rows of the matrix $G_\nu$ into the $p$-standard form with parameters $\ell_0, \ell_1, \dots, \ell_{r-1}$. This transformation has no effect on the row space of $G(D)$ and it also does not affect the row degrees $\nu_i$. We have the following upper bound on the free distance of the code.

\begin{theorem}\label{sd}
Let $G(D)= G_0 + G_1 D + \cdots + G_{\nu_1}D^{\nu_1}$ be a $p$-encoder of an $(n,k,\delta)$-convolutional code $\mathcal{C}$ in reduced form and row degrees $\nu_1 \geq \nu_2 \cdots > \nu_{k-\ell-1} = \cdots = \nu_k $ and define  $\nu=\nu_k$. Assume that the last $\ell$ rows of $G_\nu$ are in $p$-standard form with  parameters $\ell_0, \ell_1, \dots, \ell_{r-1}$. Then the free distance of $\C$ must satisfy
\begin{equation}\label{eq:01}
  d(\mathcal{C}) \leq  n(\nu+1)-(\ell_0+\ell_1+ \dots + \ell_{r-1})+1.
\end{equation}
\end{theorem}

\pf
We show that the upper bound in (\ref{eq:01}) is actually an upper bound of $d^r_0$ and therefore the result readily follows from (\ref{eq:00}). Write $G(D) = G_{0} + G_{1}D + \cdots + G_{\nu_1}D^{\nu_1}$ and denote by $G_i' $ the last $\ell$ rows of $G_i$. As matrices $G'_{\nu +1}, G'_{\nu +2},\cdots G'_{\nu_1}$ are zero, $G'(D)= G'_0 + G'_1 D + \cdots + G'_{\nu}D^{\nu}$ are the last $\ell$ rows of $G(D)$. Using that $G'_\nu$ is in the $p$-standard form, {\em i.e.}, $G'_\nu$ is equal to
{
\fontsize{7}{7}\selectfont
\arraycolsep=3pt % default: 5pt
$$
\!\!\!\!\!\! \left[ \!\!\!\!
  \begin{array}{ccccccc}
    I_{\ell_0} &  A^0_{1,0} & A^0_{2,0} & A^0_{3,0} & \cdots & A^0_{r-1,0} & A^0_{r,0} \\
    %\hdashline
    ------&  ------&------&------&------&------&------ \\
    pI_{\ell_0} &  0 & pA^0_{2,1} & pA^0_{3,1} & \cdots & pA^0_{r-1,1} & pA^0_{r,1} \\
    0 &  pI_{\ell_1} & pA^1_{2,1} & pA^1_{3,1} & \cdots & pA^1_{r-1,1} & pA^1_{r,1} \\
    ------&  ------&------&------&------&------&------ \\
    p^2I_{\ell_0} &  0 & 0 & p^2A^0_{3,2} & \cdots & p^2A^0_{r-1,2} & p^2A^0_{r,2} \\
    0 & p^2I_{\ell_1} & 0 & p^2A^1_{3,2} & \cdots & p^2A^1_{r-1,2} & p^2A^1_{r,2} \\
    0 & 0 & p^2I_{\ell_2} & p^2A^2_{3,2} & \cdots & p^2A^2_{r-1,2} & p^2A^2_{r,2} \\
    ------&------&------&------&------&------&------ \\
    \vdots&\vdots&\vdots&\vdots&\cdots&\vdots&\vdots\\
    ------&------&------&------&------&------&------ \\
    p^{r-1}I_{\ell_0} & 0 & 0 & 0 & \cdots & 0 & p^{r-1}A^0_{r,r-1} \\
    0 & p^{r-1}I_{\ell_1} & 0 & 0 & \cdots & 0 & p^{r-1}A^1_{r,r-1} \\
    0 & 0 & p^{r-1}I_{\ell_2} & 0 & \cdots & 0 & p^{r-1}A^2_{r,r-1} \\
    0 & 0 & 0 & p^{r-1}I_{\ell_3} & \cdots & 0 & p^{r-1}A^3_{r,r-1} \\
    \vdots&\vdots&\vdots&\vdots&\ddots&\vdots&\vdots\\
    0 & 0 & 0 & 0 & \cdots & p^{r-1}I_{\ell_{r-1}} & p^{r-1}A^{r-1}_{r,r-1} \\
  \end{array}
\!\!\!\! \right],
$$
}it is easy to see that the input vector $u=(0,0,\cdots, 0, 1)\in \mathcal{A}^k_p[D]$ gives a codeword $v(D)=uG(D)=u' G'(D)$ with $u'=(0,\cdots, 0, 1)\in \mathcal{A}^\ell_p[D]$. The polynomial vector $v(D)$ has the last $n - (\ell_0+\ell_1+ \cdots + \ell_{r-1}) + 1$ coordinates with weight at most $\nu +1$ and the first $\ell_0+\ell_1+ \cdots + \ell_{r-1} -1$ coordinates with weight at most $\nu$. Therefore,
\begin{eqnarray*}
  d^r_0 &\leq&  [n - (\ell_0+\ell_1+ \cdots + \ell_{r-1}) + 1](\nu + 1) + (\ell_0+\ell_1+ \cdots + \ell_{r-1} -1)\nu \\
   &=&  n(\nu+1)-(\ell_0+\ell_1+ \cdots + \ell_{r-1})+1,
\end{eqnarray*}
which concludes the proof.
\eind \pfend

\begin{remark}
We note that $\ell_0+\ell_1+ \cdots + \ell_{r-1}$ are invariants of $\C$. Indeed, if $\overline{G}(D)$ is another $p$-encoder of $\C$ in reduced form it must also have $\ell$ rows of degree $\nu$. Let $\overline{G}'(D)=\overline{G}_0' + \overline{G}'_1 D + \cdots + \overline{G}'_\nu D^\nu$ be constituted by these rows and $G'(D)=G_0' + G'_1 D + \cdots + G'_\nu D^\nu$ be as in proof of Theorem \ref{sd}. Then one can use the predictable degree property to show that $\im_{\mathcal{A}_p[D]} G'(D) = \im_{\mathcal{A}_p[D]} \overline{G}'(D) $ and furthermore $\im_{\mathcal{A}_p} G'_\nu = \im_{\mathcal{A}_p} \overline{G}'_\nu  $ which shows the claim that the $\ell_0+\ell_1+ \cdots + \ell_{r-1}$ are invariants of $\C$, see \cite{ku07} for more details.
\end{remark}

%The set of $(n,k,\delta)$-convolutional code is partitioned into sets of codes with different rows degree $\nu_1 \geq \nu_2 \cdots \geq \nu_k$.
Taking the maximum of the bound (\ref{eq:01}) over all $(n,k,\delta)$-convolutional codes we obtain the main result of \cite[Theorem 4.10]{el13}.

\begin{Corollary}   \label{free_d}
The free distance of an $(n,k,\delta)$ convolutional code $\mathcal{C}$ satisfies

\begin{equation}\label{eq:singleton_bound}
d(\mathcal{C}) \leq n\left(\left\lfloor\frac{\delta}{k}\right\rfloor+1\right)-\left\lceil \frac{k}{r}\left(\left\lfloor\frac{\delta}{k}\right\rfloor+1\right)-\frac{\delta}{r}\right\rceil+1.
\end{equation}
\end{Corollary}

\pf
Let $G(D)$ be as in Theorem \ref{sd}. The highest value of (\ref{eq:01}) is obtained by considering the maximum value of $\nu$ and the minimum value of $(\ell_0+\ell_1+ \dots + \ell_{r-1})$.
It is easy to see that the maximum value of $\nu$ is when $\nu=\left\lfloor\frac{\delta}{k}\right\rfloor$ and
$\nu_1=\nu_2= \dots=\nu_{k-\ell}=\left\lfloor\frac{\delta}{k}\right\rfloor+1$.
From this it follows that
$$
\delta=(k-\ell)\left(\left\lfloor\frac{\delta}{k}\right\rfloor+1\right)+\ell\left\lfloor\frac{\delta}{k}\right\rfloor
$$
and, thus
$$
\ell = k\left(\left\lfloor\frac{\delta}{k}\right\rfloor+1\right)-\delta.
$$
On the other hand, the values of $(\ell_0, \ell_1, \dots, \ell_{r-1})$ that minimize $\ell_0+\ell_1+ \dots + \ell_{r-1}$ and such that $\ell= \sum_{i=0}^r (r-i)\ell_i$ are the $r$-optimal set of parameters of $\ell$.  By Lemma \ref{opt_parameters}, $\ell_0+\ell_1+ \dots + \ell_{r-1}=\left\lceil\frac{\ell}{r}\right\rceil$.
Finally,
\begin{eqnarray*}
d(\mathcal{C}) & \leq & n\left(\left\lfloor\frac{\delta}{k}\right\rfloor+1\right)-\left\lceil\frac{k(\left\lfloor\frac{\delta}{k}\right\rfloor+1)-\delta}{r}\right\rceil+1 \\
& \leq &
n\left(\left\lfloor\frac{\delta}{k}\right\rfloor+1\right)-\left\lceil\frac{k}{r}\left(\left\lfloor\frac{\delta}{k}\right\rfloor+1\right)-\frac{\delta}{r}\right\rceil+1.
\end{eqnarray*}
\eind \pfend

An $(n,k,\delta)$-convolutional code over $\dubbel{Z}_{p^r}$ is said to be {\bf Maximum Distance Separable} (MDS) if
$$
d(\mathcal{C}) = n\left(\left\lfloor\frac{\delta}{k}\right\rfloor+1\right)-\left\lceil\frac{k}{r}\left(\left\lfloor\frac{\delta}{k}\right\rfloor+1\right)-
\frac{\delta}{r}\right\rceil+1.
$$

\begin{remark}
It is important to remark that the Singleton-type upper bound presented in  (\ref{eq:singleton_bound}) is derived as a corollary of the Theorem \ref{sd} by taking an $r$-optimal set parameters of $\ell = k\left(\left\lfloor\frac{\delta}{k}\right\rfloor+1\right)-\delta$ and therefore it follows that MDS convolutional codes over $\ZZ$ must have these optimal set of parameters.
\end{remark}

\section{General constructions of MDS convolutional codes over $\ZZ$}

In this section we present a general procedure for building (non necessarily free) MDS convolutional codes over $\ZZ$. The idea is to start from well-known constructions of MDS convolutional codes over $\Z_p$ and then \emph{lift} them to $\ZZ$ in such a way that the resulting convolutional code is MDS over $\ZZ$. This method is direct and works for any given set of parameters $(n,k,\delta)$.  \\

For the sake of simplicity of exposition, we first assume that $k \mid \delta$ and consequently the row degrees of any $p$-encoder $G(D)$ of $\C$ are $\nu=\nu_1 =\cdots = \nu_k = \frac{\delta}{k}$ and thus $\ell=k$. The general case will be treated at the end of the section. Hence, the MDS $(n,k,\delta)$-convolutional $\C$ that we aim to construct must satisfy
$$
d(\mathcal{C}) = n\left(\left\lfloor\frac{\delta}{k}\right\rfloor+1\right)-\left\lceil\frac{k}{r}\left(\left\lfloor\frac{\delta}{k}\right\rfloor+1\right)-
\frac{\delta}{r}\right\rceil+1.
$$
Note that
$$
n\left(\left\lfloor\frac{\delta}{k}\right\rfloor+1\right)-\left\lceil\frac{k}{r}\left(\left\lfloor\frac{\delta}{k}\right\rfloor+1\right)-
\frac{\delta}{r}\right\rceil+1 = n(\nu+1)-(k_0+k_1+ \dots + k_{r-1})+1
$$
where $k_0, k_1, \dots, k_{r-1}$ is an $r$-optimal set of parameters of $k$.

Take $(\widetilde{k}=k_0+k_1+ \dots+ k_{r-1})$ and $\widetilde{\delta}=\nu\widetilde{k}$, and let us consider any of the well-known construction of MDS convolutional codes $\widetilde{\mathcal{C}}$ (see \cite{gu14,NaRo2015,sm01a}) with length $n$, dimension $\widetilde{k}$ and degree $\widetilde{\delta}$ over a field $\dubbel{Z}_{p}$.\\

The distance of $\widetilde{\mathcal{C}}$ equals (see \cite{ro99a1})
$$
d(\widetilde{\mathcal{C}})=(n-\widetilde{k})\left(\left\lfloor\frac{\widetilde{\delta}}{\widetilde{k}}\right\rfloor+1\right)+\widetilde{\delta}+1.
$$
Let
\begin{equation}\label{eq:05}
\widetilde{G}(D)=
\left[
  \begin{array}{c}
    \widetilde{G}_{k_0}(D) \\
    ---- \\
    \widetilde{G}_{k_1}(D) \\
    ---- \\
     \vdots \\
     ---- \\
     \widetilde{G}_{k_{r-1}}(D)
  \end{array}
\right]\in \dubbel{Z}_{p}[D]^{\widetilde{k} \times n}
\end{equation}
be an encoder of $\widetilde{\mathcal{C}}$ in reduced form, where $\widetilde{G}_{k_i}(D)$ is a $k_i \times n$ matrix, $i=0,1, \dots, r-1$, .

By Lemma \ref{opt_parameters}, $\widetilde{k}=\left\lceil\frac{k}{r}\right \rceil$  and since $\widetilde{\delta}=\nu\widetilde{k}$ we get that
\begin{equation}\label{eq:04}
d(\widetilde{\mathcal{C}})=n\left(\nu+1\right)-\left\lceil\frac{k}{r}\right \rceil+1.
\end{equation}

Next, we \emph{lift} $\widetilde{G}(D)$ to construct a $k \times n$ matrix $G(D)$ in $\ZZ$ as follows,
\begin{equation}\label{eq:02}
G(D)=
\left[
  \begin{array}{c}
    \widetilde{G}_{k_0}(D) \\
     p\widetilde{G}_{k_0}(D) \\
      \vdots \\
       p^{r-1}\widetilde{G}_{k_0}(D) \\
       ---- \\
        p\widetilde{G}_{k_1}(D) \\
         p^2\widetilde{G}_{k_1}(D) \\
          \vdots \\
           p^{r-1}\widetilde{G}_{k_1}(D) \\
           ---- \\
            \vdots \\
             ---- \\
             p^{r-1}\widetilde{G}_{k_{r-1}}(D)
  \end{array}
\right].
\end{equation}
%with $rl_0+(r-1)l_1+ \dots+l_{r-1}=k$ and we prove the next Lemma.

\begin{Lemma}\label{lem:00}
The matrix $G(D)$ defined in (\ref{eq:02}) is a $p$-encoder in reduced form with row degrees all equal to $\nu$. Moreover, the convolutional code generated by $G(D)$ has $p$-dimension $k$ and $p$-degree $\delta$.
\end{Lemma}

\pf
Since all the rows of $\widetilde{G}(D)$ have row degrees $\nu$,
the rows of $G(D)$ have also degree $\nu$. From the construction of $G(D)$ it is straightforward to verify that its rows form a $p$-generator sequence.
It remains to show that $G(D)$ is in reduced form, \emph{i.e.}, the rows of
$$
G^{lc}=
\left[
  \begin{array}{c}
    \widetilde{G}_{k_0}^{lc} \\
     p\widetilde{G}_{k_0}^{lc} \\
      \vdots \\
       p^{r-1}\widetilde{G}_{k_0}^{lc} \\
       ---- \\
        p\widetilde{G}_{k_1}^{lc} \\
         p^2\widetilde{G}_{k_1}^{lc} \\
          \vdots \\
           p^{r-1}\widetilde{G}_{k_1}^{lc} \\
           ---- \\
            \vdots \\
             ---- \\
             p^{r-1}\widetilde{G}_{k_{r-1}}^{lc} \\
  \end{array}
\right],
$$
are $p$-linearly independent.
This amounts to show that for $a_j^i \in \mathcal{A}_p$, with $i=j, \dots, r-1$ and $j=0,\dots, r-1$,
\begin{equation}\label{eq:03}
\begin{split}
  a_0^0 \widetilde{G}_{k_0}^{lc}+ a_0^1 p\widetilde{G}_{k_0}^{lc} + \cdots + & a_0^{r-1} p^{r-1}\widetilde{G}_{k_0}^{lc} + a_1^1 p\widetilde{G}_{k_1}^{lc}+ a_1^2 p^2\widetilde{G}_{k_1}^{lc}+ \cdots + \\
   & + \cdots + a_1^{r-1} p^{r-1}\widetilde{G}_{k_1}^{lc} + \cdots + a_{r-1}^{r-1} p^{r-1}\widetilde{G}_{k_{r-1}}^{lc}=0
\end{split}
\end{equation}
  implies that  $a_0^0=a_0^1=\dots=a_0^{r-1}=a_1^1=a_1^1=\dots=a_1^{r-1}=\dots=a_{r-1}^{r-1}=0.$

Note that, multiplying (\ref{eq:03}) by $p^{r-1}$
%\begin{description}
%   \item[ ] $\widetilde{G}_{l_0}^{lc}a_0^0 + p\widetilde{G}_{l_0}^{lc}a_0^1 + \cdots + p^{r-1}\widetilde{G}_{l_0}^{lc}a_0^{r-1} + p\widetilde{G}_{l_1}^{lc}a_1^1 + p^2\widetilde{G}_{l_1}^{lc}a_1^2 + \cdots + p^{r-1}\widetilde{G}_{l_1}^{lc}a_1^{r-1} + \cdots + p^{r-1}\widetilde{G}_{l_{r-1}}^{lc}a_{r-1}^{r-1}=0$\\
% \end{description}
we obtain $a_0^0  p^{r-1}\widetilde{G}_{k_0}^{lc} =0$. As $\widetilde{G}(D)$ is in reduced form, $\widetilde{G}_{k_0}^{lc}$ must be full row rank over $\Z_p$ and therefore $a_0^0 p^{r-1}\widetilde{G}_{k_0}^{lc} =0$ implies $a_0^0=0$. Proceeding in the same way,
by successively multiplying (\ref{eq:03}) by $p^{r-2}, \dots, 1$,
we show that $a_j^i=0$, with $i=j, \dots, r-1$ and $j=0,\dots, r-1$.

For the proof of the last statement note that since $\widetilde{k}=k_0 + k_1 + \cdots + k_{r-1}$ and $(k_0, \dots, k_{r-1})$ is an $r$-optimal set of parameters of $k$ we obtain that $G(D)$ has $k$ rows, \emph{i.e.}, $\C$ has $p$-dimension equal to $k$.
Moreover, the degree of $\mathcal{C}$ is $\nu k = \frac{\delta}{k} k= \delta$.
\eind \pfend

%
%
%Let $\mathcal{C}$ be an $(n,k,\delta)$-convolutional code with $p$-encoder $G(D)$ as defined above.\\
%It is easy to see that
%{
%\fontsize{7}{7}\selectfont
%\arraycolsep=3pt % default: 5pt
%$$
%\hat{G}(D)=
%\left[
%  \begin{array}{c|cc|ccc|c|cccc}
%    \widetilde{G}_{l_0}(D) & p\widetilde{G}_{l_0}(D) & p\widetilde{G}_{l_1}(D) & p^2\widetilde{G}_{l_0}(D) & p^2\widetilde{G}_{l_1}(D) & p^2\widetilde{G}_{l_2}(D) & \cdots & p^{r-1}\widetilde{G}_{l_0}(D) & p^{r-1}\widetilde{G}_{l_1}(D) & \cdots & p^{r-1}\widetilde{G}_{l_{r-1}}(D) \\
%  \end{array}
%\right]
%$$
%}
%is still a reduced $p$-encoder of $\mathcal{C}$.

The following technical lemma will be used in the next theorem. First, we need to define the order of a codeword. If $v(D)\in \ZZ[D]\setminus \{ 0 \}$ we define the \emph{order} of $v(D)$, denoted by $ord(v(D))$, as the $j\in \{ 1,2,\dots, r \}$ such that $p^j v(D)=0$ and $p^{j-1} v(D)\neq 0$.

\begin{Lemma}\label{mds}
Let $\mathcal{C}$ be the convolutional code generated by the $p$-encoder $G(D)$  defined in (\ref{eq:02}) and $\widetilde{G}(D)$ be as in (\ref{eq:05}). Then, if $v(D) \in \C$ has order $j$,
\begin{equation*}
  p^{r-j}v(D) \in \mbox{ \em Im}_{\mathcal{A}_{p}[D]} \ p^{r-1} \widetilde{G}(D).
\end{equation*}
\end{Lemma}

\pf
Since the matrix $\widetilde{G}(D)$ defined in (\ref{eq:05}) is full row rank over $\dubbel{Z}_p[D]$, it follows that, for any nonzero codeword of $\mathcal{C}$, $v(D)=\sum_{i=0}^{r-1} {\sum_{l=i}^{r-1}{u_i^l(D) p^j \widetilde{G}_{k_i}}}$, with $u_i^l(D) \in \mathcal{A}_p^{k_i}[D]$,
\begin{equation}\label{eq:06}
ord(v(D))= \max\limits_{i,l:u_i^l(D)\neq 0}{ord(p^l\widetilde{G}_{k_i})}.
\end{equation}
Thus, if $v(D)$ has order $j$ then $p^{r-j}v(D)$ has order one and therefore, by (\ref{eq:06}), $p^{r-j}v(D) \in \im_{\mathcal{A}_{p}[D]} \ p^{r-1} \widetilde{G}(D)$.
\eind \pfend

Now we are ready to present the result that shows that our construction is indeed MDS.

\begin{Thm} \label{thm}
Let $\mathcal{C}$ be the $(n,k,\delta)$-convolutional code with $k \mid \delta$ and $p$-encoder $G(D)$ as in (\ref{eq:02}). Then, $\C$ is MDS, \emph{i.e.},

\begin{equation*}
\begin{split}
d(\mathcal{C}) =& n\left(\left\lfloor\frac{\delta}{k}\right\rfloor+1\right)-\left\lceil \frac{k}{r}\left(\left\lfloor\frac{\delta}{k}\right\rfloor+1\right)-\frac{\delta}{r}\right\rceil+1 
%= & n \left( \frac{\delta}{k}+1 \right)-\left\lceil\frac{k}{r}\right \rceil+1.
\end{split}
\end{equation*}
\end{Thm}

\pf
Since $k \mid \delta$ we can simplify the formula of the distance of $\C$, namely, 
$$
n\left(\left\lfloor\frac{\delta}{k}\right\rfloor+1\right)-\left\lceil \frac{k}{r}\left(\left\lfloor\frac{\delta}{k}\right\rfloor+1\right)-\frac{\delta}{r}\right\rceil+1
=  n \left( \frac{\delta}{k}+1 \right)-\left\lceil\frac{k}{r}\right \rceil+1.
$$
Let $v(D) \in \mathcal{C}\setminus \{0\}$. Obviously, $wt(v(D))\geq wt(p^{r-j}v(D))$, where $j$ is the order of $v(D)$.
By Lemma \ref{mds},
$$
 wt(p^{r-j}v(D))=wt(p^{r-1}u(D)\widetilde{G}(D)),
$$
for some $u(D) \in \mathcal{A}_p^k[D]$.

Note that, since $u(D) \in \mathcal{A}_p^k[D]$,
$$
wt(p^{r-1}u(D)\widetilde{G}(D))=wt_p(\bar{u}(D)\widetilde{G}(D)),
$$
where $\bar{u}(D)=u(D)$ is the projection of $u(D)$ over $\dubbel{Z}_p[D]$ and $wt_p$ represents the Hamming weight over $\dubbel{Z}_p$.
This together with the fact that $\widetilde{\C}$ is an MDS convolutional code over $\Z_p$ shows that
$$
wt(p^{r-1}u(D)\widetilde{G}(D))\geq (n-\widetilde{k})\left(\left\lfloor\frac{\widetilde{\delta}}{\widetilde{k}}\right\rfloor+1\right)+\widetilde{\delta}+1.
$$
It is straightforward to check that for $\widetilde{\delta}= \nu\widetilde{k}= \frac{\delta}{k}\widetilde{k}$ and $\widetilde{k}= \left \lceil\frac{k}{r} \right\rceil$ this lower bound coincides with the upper-bound given in Corolary \ref{free_d}.
This show that $d(\mathcal{C})=n \left( \frac{\delta}{k}+1 \right)-\left\lceil\frac{k}{r}\right \rceil+1$.
\eind \pfend

Let us now assume that $k \nmid \delta$. In this case we know that
$\ell = k\left(\left\lfloor\frac{\delta}{k}\right\rfloor+1\right)-\delta$,
and we select $(\ell_0, \dots, \ell_{r-1})$ to be an $r$-optimal set of parameters of $\ell$ in order to construct
an MDS $(n,k,\delta)$-convolutional $\C$, \emph{i.e.}, such that 
$$
d(\mathcal{C}) =  n(\nu+1)-(\ell_0+\ell_1+ \dots + \ell_{r-1})+1.
$$

Let $a,b \in \mathbb{N}_0$ such that $k-\ell=ar+b$, with $b<r$. Take $\widetilde{k}=a+1+\ell_0+\ell_1+ \dots+ \ell_{r-1}$, $\nu=\left\lfloor\frac{\delta}{k}\right\rfloor$ and $\widetilde{\delta}=(a+1)(\nu+1)+(\ell_0+\ell_1+ \dots+ \ell_{r-1})\nu$, and let
$\widetilde{\mathcal{C}}$ be an MDS convolutional code of length $n$, dimension $\widetilde{k}$ and degree $\widetilde{\delta}$ over a field $\dubbel{Z}_{p}$.
Construct
$$
\widetilde{G}(D)=
\left[
  \begin{array}{c}
    \widetilde{G}_{a}(D) \\
    ---- \\
    \widetilde{G}_{1}(D) \\
    ---- \\
    \widetilde{G}_{\ell_0}(D) \\
    ---- \\
    \widetilde{G}_{\ell_1}(D) \\
    ---- \\
     \vdots \\
     ---- \\
     \widetilde{G}_{\ell_{r-1}}(D)
  \end{array}
\right]\in \dubbel{Z}_{p}[D]^{\widetilde{k} \times n}
$$
to be an encoder of $\widetilde{\mathcal{C}}$ in reduced form, where $\widetilde{G}_{a}(D)$ is a $a \times n$ matrix,
$\widetilde{G}_{1}(D)$ is a $1 \times n$ matrix with row degrees $\nu + 1$ and $\widetilde{G}_{\ell_i}(D)$ is an $\ell_i \times n$ matrix with row degrees $\nu$, $i=0,1, \dots, r-1$.

Since $\widetilde{\mathcal{C}}$  is an MDS $(n,\widetilde{k},\widetilde{\delta})$-convolutional code over $\dubbel{Z}_p $,
the distance equals (see \cite{ro99a1})
$$
d(\widetilde{\mathcal{C}})=(n-\widetilde{k})\left(\left\lfloor\frac{\widetilde{\delta}}{\widetilde{k}}\right\rfloor+1\right)+\widetilde{\delta}+1.
$$
Note that from $\widetilde{k}=a+1+\ell_0+\ell_1+ \dots+ \ell_{r-1}$ and $\widetilde{\delta}=(a+1)(\nu+1)+(\ell_0+\ell_1+ \dots+ \ell_{r-1})\nu$ we have that $\frac{\widetilde{\delta}}{\widetilde{k}}= \nu+\frac{a+1}{\widetilde{k}}$, and therefore $\nu=\left\lfloor\frac{\widetilde{\delta}}{\widetilde{k}}\right \rfloor$, and also that
$$
d(\widetilde{\mathcal{C}})= n(\nu+1)-(\ell_0+\ell_1+ \dots + \ell_{r-1})+1.
$$
Now, let us consider the following $k \times n$ matrix in $\ZZ$,
$$
G(D)=
\left[
  \begin{array}{c}
   \widetilde{G}_{a}(D) \\
     p\widetilde{G}_{a}(D) \\
      \vdots \\
       p^{r-1}\widetilde{G}_{a}(D) \\
       ---- \\
     p^{r-b}\widetilde{G}_{1}(D) \\
      \vdots \\
       p^{r-1}\widetilde{G}_{1}(D) \\
       ---- \\
    \widetilde{G}_{\ell_0}(D) \\
     p\widetilde{G}_{\ell_0}(D) \\
      \vdots \\
       p^{r-1}\widetilde{G}_{\ell_0}(D) \\
       ---- \\
        p\widetilde{G}_{\ell_1}(D) \\
         p^2\widetilde{G}_{\ell_1}(D) \\
          \vdots \\
           p^{r-1}\widetilde{G}_{\ell_1}(D) \\
           ---- \\
            \vdots \\
             ---- \\
             p^{r-1}\widetilde{G}_{v_{r-1}}(D)
  \end{array}
\right].
$$
Applying the same reasoning as in the proofs of Lemma \ref{lem:00} and Theorem \ref{thm}, we conclude that $G(D)$ is a $p$-encoder in reduced form of an MDS $(n,k,\delta)$-convolutional code.

\section{Conclusions}

In this paper we further investigated distance properties of convolutional codes over $\Z_{p^r}$ extending the results presented in \cite{no01} for block codes. In particular we have focused our attention on a generalization of the Singleton bound and the class of MDS codes. Continuing the work in \cite{el13} we presented novel results in this direction, \emph{e.g.}, a method to construct MDS for any set of given parameters.

It will be interesting to study these codes equipped with a homogeneous weight \cite{gr00} or different metrics. Another interesting avenue of research is to investigate how we can use these results to construct non-linear (binary) trellis codes.

\section{Acknowledgements}

This work was supported by Portuguese funds through the CIDMA - Center for Research and Development in Mathematics and Applications, and the Portuguese Foundation for Science and Technology (FCT-Funda\c{c}\~ao para a Ci\^encia e a Tecnologia), within project PEst-UID/MAT/04106/2013.

%\begin{acknowledgements}
%\end{acknowledgements}

\bibliographystyle{spmpsci}      % mathematics and physical sciences

\bibliography{biblio}   % name your BibTeX data base

%% Non-BibTeX users please use
%\begin{thebibliography}{}
%%
%% and use \bibitem to create references. Consult the Instructions
%% for authors for reference list style.
%\end{thebibliography}

\end{document}